\documentclass[12pt]{article} 

\usepackage{epsfig} 
\usepackage{amsfonts}
\usepackage{amssymb}
\usepackage{amsmath}  
\usepackage{amsbsy} 
 
\newlength{\largfig}
\def\beq{\begin{equation}} 
\def\eeq{\end{equation}} 
 
\def\bea{\begin{eqnarray}} 
\def\eea{\end{eqnarray}} 
   
\def\mc{\mathcal}
\def\mr{\mathrm}

\def \eps{\epsilon}

\def\({\left(} 
\def\){\right)}

\def\sla#1{\ifmmode% 
\setbox0=\hbox{$#1$}% 
\setbox1=\hbox to\wd0{\hss$/$\hss}\else% 
\setbox0=\hbox{#1}% 
\setbox1=\hbox to\wd0{\hss/\hss}\fi% 
#1\hskip-\wd0\box1 }

\def\non{\nonumber}

\def\al{\alpha}
\def\be{\beta}
\def\ga{\gamma}
\def\lam{\lambda}
\def\Ga{\Gamma}

\def\tpii{2\pi i}

\def\fpi{4 \pi}

\def\hs{\hspace*}

\newcommand{\ca}{C_A}
\newcommand{\cf}{C_F}

\newskip\humongous \humongous=0pt plus 1000pt minus 1000pt

\newif\ifdtup

\def\baselinestretch{1.1} 
\catcode`@=12 
 
\catcode`\@=11 
 
\@addtoreset{equation}{section} 

\def\theequation{\thesection.\arabic{equation}} 
 
\def\@normalsize{\@setsize\normalsize{15pt}\xiipt\@xiipt 
\abovedisplayskip 14pt plus3pt minus3pt% 
\belowdisplayskip \abovedisplayskip 
\abovedisplayshortskip \z@ plus3pt% 
\belowdisplayshortskip 7pt plus3.5pt minus0pt} 
 
\def\small{\@setsize\small{13.6pt}\xipt\@xipt 
\abovedisplayskip 13pt plus3pt minus3pt% 
\belowdisplayskip \abovedisplayskip 
\abovedisplayshortskip \z@ plus3pt% 
\belowdisplayshortskip 7pt plus3.5pt minus0pt 
\def\@listi{\parsep 4.5pt plus 2pt minus 1pt 
     \itemsep \parsep 
     \topsep 9pt plus 3pt minus 3pt}} 
 
\@twosidetrue

\catcode`\@=11  
\def\section{\@startsection{section}{1}{\z@}{3.5ex plus 1ex minus 
   .2ex}{2.3ex plus .2ex}{\large\bf}}

\def\thesection{\arabic{section}} 
\def\thesubsection{\arabic{section}.\arabic{subsection}} 
\def\thesubsubsection{\arabic{section}.\arabic{subsection}.\arabic{subsubsection}} 
 
\def\appendix{\setcounter{section}{0} 
 \def\thesection{\Alph{section}} 
 \def\theequation{\Alph{section}.\arabic{equation}} 
\def\thesubsection{\Alph{section}.\arabic{subsection}} 
\def\thesubsubsection{\Alph{section}.\arabic{subsection}.\arabic{subsubsection}} 
 
\def\section{\@startsection{section}{1}{\z@}{3.5ex plus 1ex minus 
   .2ex}{2.3ex plus .2ex}{\large\bf}} 
}

\def \eps{\epsilon} 
\def \to   {\mbox{$\rightarrow$}}

\newcount\minutes 
\newcount\scratch 
 
\def\timestamp{% 
\scratch=\time 
\divide\scratch by 60 
\edef\hours{\the\scratch} 
\multiply\scratch by 60 
\minutes=\time 
\advance\minutes by -\scratch 
---$\,$\hours:\null 
\ifnum\minutes< 10 0\fi 
\the\minutes}

\begin{document} 
\begin{titlepage} 
\nopagebreak 
{\flushright{ 
        \begin{minipage}{2.8cm} 
         KEK-TH-1222 \\ 
        \end{minipage}        } 
 
} 
\vfill 
\begin{center} 
{\LARGE \bf \sc 
 \baselineskip 0.9cm 
 \parskip 0.9cm 
Mixed QCD-electroweak contributions to \\[2mm]
Higgs-plus-dijet production at the LHC
} 
\vskip 0.5cm  
{\large   
A.\  Bredenstein$^1$, K.\  Hagiwara$^{1,2}$, B.\ J\"ager$^1$
}   
\vskip .2cm  
{$^1$ {\it KEK Theory Division, Tsukuba 305-0801, Japan}}\\
{$^2$ {\it The Graduate University of Advanced Studies (Sokendai), \\
	   Tsukuba 305-0801, Japan}}\\
 
 \vskip 
1.3cm     
\end{center} 
 
\nopagebreak 
%\vfill 
%\vskip 3cm 
\begin{abstract}
We present a calculation of interference effects in $Hjj$ production via gluon
fusion and via vector boson fusion, respectively, beyond tree level. We
reproduce results recently discussed in the literature, but go beyond this
calculation by including a class of diagrams not considered previously. Special
care is taken in developing a numerically stable and flexible parton level Monte-Carlo program 
which allows us to study cross sections and kinematic
distributions within experimentally relevant selection cuts. 
Loop-induced interference contributions are found to exhibit kinematical
distributions different in shape from vector boson fusion. Due to
the small interference cross section and cancelation among different quark flavor contributions 
their impact on the signal process is found
to be negligible in all regions of phase space, however. 
\end{abstract} 
\vfill 
\today \timestamp \hfill 
\vfill 

% PACS: 14.80.Bn
\end{titlepage} 
\newpage               
%
%%%%%%%%%%%%%%%%%%%%%%%%%%%%%%%%%%%%%%%%%%%%%%%%%%%%%%%%%%%%%%%%%
%%%%%%%%%%%%%%%%%%%%%%%%%%%%%%%%%%%%%%%%%%%%%%%%%%%%%%%%%%%%%%%%%
%
\section{Introduction}
\label{sec:intro}
Higgs production via weak boson fusion (WBF), i.e., the reaction $pp\to Hjj$,
mediated by $t$-channel weak boson exchange, constitutes a particularly
promising production mechanism for the Higgs boson.  Due to the distinctive 
signature of two hard jets accompanying the decay products of the Higgs boson,
this channel is discussed as possible discovery mode for a scalar, CP-even boson
as predicted by the Standard Model (SM) \cite{ATLAS-CMS,VBF:H}, 
and as powerful tool for a later
determination of its couplings \cite{VBF:C}. Furthermore, WBF could 
be employed in studying
deviations from the SM expectations and help to spot signatures of physics
beyond the standard model. This can only be achieved, however, if accurate 
measurements are matched by precision calculations of SM signal and background
processes and predictions for possible new physics scenarios. 

Next-to-leading order (NLO) QCD corrections to the SM $pp\to Hjj$ WBF signal 
are available for cross sections \cite{HVW:Hjj} and distributions
\cite{NLO:Hjj}. At the same level of accuracy, some of the most important
background processes such as $Vjj$~\cite{OZ:Vjj} and $VVjj$~\cite{JOZ:VV} 
production in WBF, $t\bar t$~\cite{tt} and $t\bar tj$~\cite{DUW:ttj} production 
are known. Beyond the standard model, Monte Carlo studies have 
been performed for WBF $Hjj$
production in the presence of anomalous gauge boson couplings \cite{VBF:CP} and in
the context of supersymmetric models \cite{susy-wbf}. 
Recently, NLO electroweak (EW) corrections for cross sections and distributions
have been presented~\cite{CDD:Hjj}. Finite parts of the NNLO-QCD corrections
have been calculated in \cite{NNLO:VBF} and found to be negligible.

An irreducible background to the Higgs signal in WBF is constituted by $Hjj$
production via gluon fusion (GF). Higgs production via GF is mediated by a heavy
quark loop. An exact calculation of  $pp\to Hjj$ via GF at the lowest
non-vanishing order has been performed in \cite{GF:LO}. NLO-QCD calculations
have made use of the large top mass limit, where the coupling of the Higgs to
gluons is parameterized by an effective vertex \cite{GF:NLO}. Phenomenological
studies have revealed the complementary features of the WBF and the GF $Hjj$
production processes, suggesting search strategies for suppressing the GF
channel as background to the clean WBF signature. 
More recently, GF has also been considered as a signal process \cite{GF:sig}, 
because the dijet angular correlation is sensitive to the CP parity of the Higgs
boson. 

Although GF and WBF are usually considered as separate reactions, 
their interference in the $qq\to qqH$  subprocesses is possible. 
In Refs.~\cite{cg:dip,as:int}
interference effects at tree-level due to identical flavor effects have
explicitly been shown to be tiny and entirely negligible for cross sections and
distributions. The authors of \cite{as:int} speculated that loop-induced
interference effects should be large. An explicit calculation revealed, however,
that loop-induced interference effects are also small \cite{abhs:int}. 

We present a similar calculation for the same process, studying interference effects between GF and
WBF which emerge beyond tree level. Being performed with entirely different
methods, our work  confirms the main findings of
Ref.~\cite{abhs:int} and extends it in two respects:  
First, we include a class of real emission contributions which has 
been neglected in \cite{abhs:int}. 
Second, we develop a fully flexible parton level Monte Carlo program 
which allows us to study cross sections as well as arbitrary distributions 
within experimentally feasible selection cuts. Being implemented in the modular
{\tt vbfnlo} environment \cite{vbfnlo}, the impact of the interference contributions on the
WBF signal is studied in detail. 

We give a thorough outline of our calculation in Sec.~\ref{sec:calc}. 
Numerical results are discussed in Sec.~\ref{sec:res}. We conclude with a
brief summary in Sec.~\ref{sec:concl}. 

%
%%%%%%%%%%%%%%%%%%%%%%%%%%%%%%%%%%%%%%%%%%%%%%%%%%%%%%%%%%%%%%%%%
%

\section{Elements of the calculation}
\label{sec:calc}
\subsection{General framework}
\begin{figure}[!tb] 
\centerline{ 
\epsfig{figure=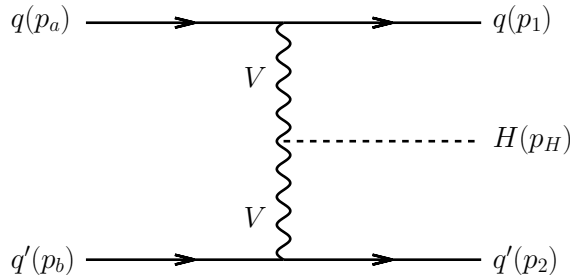,width=0.45\textwidth,clip=}
} 
\caption
{\label{fig:vbf-lo} 
Feynman diagram for the tree level process $qq'\to qq'H$ 
via WBF. 
}
\end{figure}

$Hjj$ production in WBF mainly proceeds via quark scattering, $qq\to qqH$.  
In Ref.~\cite{cg:dip} contributions from identical-flavor annihilation
processes such as $q\bar q\to Z^\star\to ZH$ with subsequent decay $Z\to q\bar
q$ or similar $WH$ production channels have been shown to be entirely negligible
in the phase-space regions where WBF can be observed experimentally. In
the same work,  identical quark interference effects from $qq\to qqH$ and
crossing-related channels were demonstrated to 
affect cross sections and kinematic distributions at an insignificant level.
This finding was confirmed by Ref.~\cite{as:int}. 
In the following we will therefore restrict our discussion to quark scattering
via exchange of a weak boson in the $t$-channel, i.e.\ the reaction $qq'\to
qq'H$, where $q$ and $q'$ stand for quarks of different flavor, 
see Fig.~\ref{fig:vbf-lo}. We will refer to the respective
tree level scattering amplitude by
$\mc{M}_{\rm{WBF}}^{(0)}$. Color factors are not included in
$\mc{M}_{\rm{WBF}}^{(0)}$ and will be denoted separately. 
Adaptation for the crossed processes 
$q\bar q'\to q\bar q'H$, $\bar q q'\to \bar q q' H$, and  $\bar q\bar q'\to\bar q\bar q'
H$ is straightforward. 
% 
%%%%%%%%%%%%%%%%%%%%%%%%%%%%%%%%%%%%%%%%%%%%%%%%%%%%%%%%%%%%%%%%%%%%%%
%
% 
%%%%%%%%%%%%%%%%%%%%%%%%%%%%%%%%%%%%%%%%%%%%%%%%%%%%%%%%%%%%%%%%%%%%%%

\begin{figure}[!tb] 
\centerline{ 
\epsfig{figure=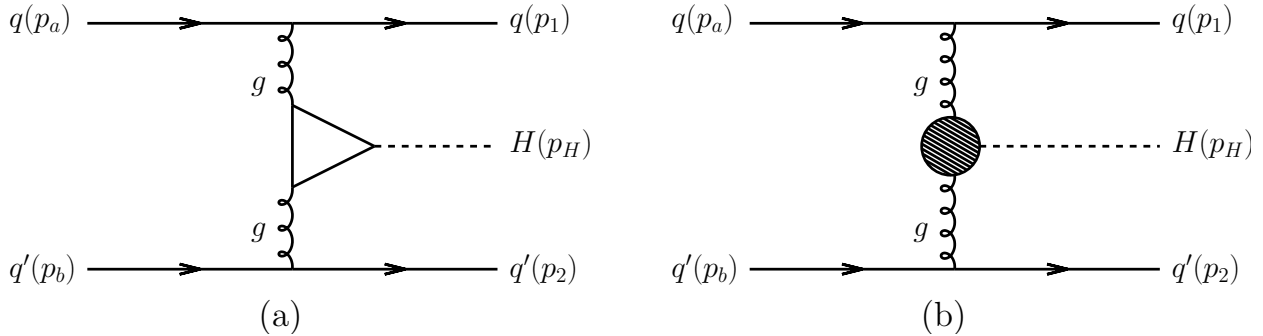,width=\textwidth,clip=}
} 
\caption
{\label{fig:gf-lo} 
Feynman diagrams for the tree level process $qq'\to qq'H$ 
via GF, mediated (a) by a top-quark loop and (b) by the effective $Hgg$ vertex
of Eq.~(\ref{eq:hgg-eff}). 
}
\end{figure}

Higgs production in quark scattering reactions,  mediated by
a gluon in the $t$-channel which couples to the Higgs boson via a
top-quark loop is depicted in Fig.~\ref{fig:gf-lo}(a). 
% 
%%%%%%%%%%%%%%%%%%%%%%%%%%%%%%%%%%%%%%%%%%%%%%%%%%%%%%%%%%%%%%%%%%%%%%
%
% 
%%%%%%%%%%%%%%%%%%%%%%%%%%%%%%%%%%%%%%%%%%%%%%%%%%%%%%%%%%%%%%%%%%%%%%
%
%
For a Higgs mass well below the top-pair production threshold, 
the coupling of the gluon to a scalar, CP-even Higgs boson can
be parameterized by an effective Lagrangian of the form
\beq
\label{eq:hgg-eff}
\mc{L}_{eff} = \frac{\alpha_s}{12\pi v} H G_{\mu\nu}^a G^{a\,\mu\nu}\, ,
\eeq 
where $G$ denotes the gluonic field tensor and $v=246$~GeV the vacuum expectation value
of the Higgs boson. The respective Feynman diagram is depicted in
Fig.~\ref{fig:gf-lo}(b). Throughout this work we will employ this effective
coupling for the $Hgg$ vertex. 
In the following, we denote the lowest order scattering
amplitude for $qq'\to qq'H$ via GF by  $\mc{M}_{\rm{GF}}^{(0)}$. Analogous to the WBF
case,  color
factors are not included in the amplitude  $\mc{M}_{\rm{GF}}^{(0)}$.

\begin{figure}[!tb] 
\centerline{ 
\epsfig{figure=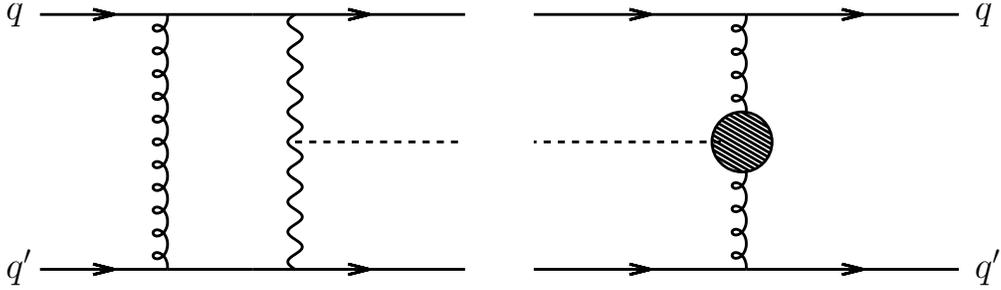,width=0.8\textwidth,clip=}
} 
\caption
{\label{fig:vbf-loop} 
Representative loop contribution to the interference cross section for $qq'\to qq'H$ via WBF 
and GF, respectively, at  order $\mc{O}(\alpha^2 \alpha_s^3)$, 
where the 1-loop WBF amplitude interferes with the tree-level GF amplitude.
}
\end{figure}

\begin{figure}[!tb] 
\centerline{ 
\epsfig{figure=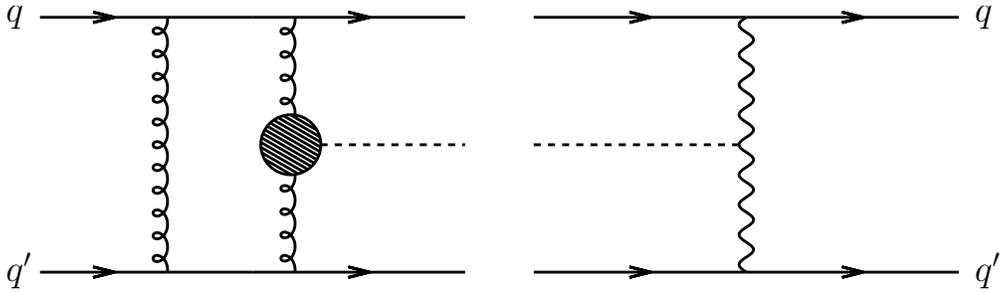,width=0.8\textwidth,clip=}
} 
\caption
{\label{fig:gf-loop} 
Representative loop contribution to the interference cross section for 
$qq'\to qq'H$ via WBF 
and GF, respectively, at  order $\mc{O}(\alpha^2 \alpha_s^3)$, where the 
1-loop GF amplitude interferes with the tree-level WBF amplitude.
}
\end{figure}

At tree level, the GF and WBF production channels for $qq'\to qq'H$ do not
interfere due to the color structure of the two processes. 
An interference between GF and the {\em neutral-current} contributions to 
WBF becomes possible, however, if an additional gluon emission
is considered. Flavor-changing $WW$-fusion diagrams cannot interfere with 
the flavor-conserving gluon exchange diagrams. 
For the neutral-current mode, two types of loop contributions emerge:
\begin{itemize}
\item[(1)]
One-loop diagrams,   
where a gluon is exchanged between the upper and the lower fermion
line in the WBF diagram of Fig.~\ref{fig:vbf-lo} (for $V=Z$). 
The respective loop amplitude $\mc{M}_{\rm{WBF}}^{(1-\rm{loop})}$ yields
non-vanishing contributions at order  $\mc{O}(\alpha^2 \alpha_s^3)$ 
when interfering 
with the tree-level GF production amplitude  $\mc{M}_{\rm{GF}}^{(0)}$. 
Here, we count the $HZZ$ coupling as $\alpha v$ and the $Hgg$ coupling as
$\alpha_s/v$; 
see Fig.~\ref{fig:vbf-loop} for a representative Feynman graph. 

\item[(2)]
GF diagrams  with an extra gluon
exchanged between the upper and the lower quark line,  
$\mc{M}_{\rm{GF}}^{(1-\rm{loop})}$, also 
contribute at order $\mc{O}(\alpha^2 \alpha_s^3)$
when interfering  with the tree-level $ZZ$-fusion  amplitude 
$\mc{M}_{\rm{WBF}}^{(0)}$ as depicted in  Fig.~\ref{fig:gf-loop}. 
\end{itemize}
All relevant loop diagrams involve pentagon diagrams. Box, triangle, and
bubble diagrams do not contribute due to color conservation. 
%
% 
%%%%%%%%%%%%%%%%%%%%%%%%%%%%%%%%%%%%%%%%%%%%%%%%%%%%%%%%%%%%%%%%%%%%%%
%
% 
%%%%%%%%%%%%%%%%%%%%%%%%%%%%%%%%%%%%%%%%%%%%%%%%%%%%%%%%%%%%%%%%%%%%%%
%
% 
%%%%%%%%%%%%%%%%%%%%%%%%%%%%%%%%%%%%%%%%%%%%%%%%%%%%%%%%%%%%%%%%%%%%%%
%
% 
%%%%%%%%%%%%%%%%%%%%%%%%%%%%%%%%%%%%%%%%%%%%%%%%%%%%%%%%%%%%%%%%%%%%%%
%
%

\begin{figure}[!tb] 
\centerline{ 
\epsfig{figure=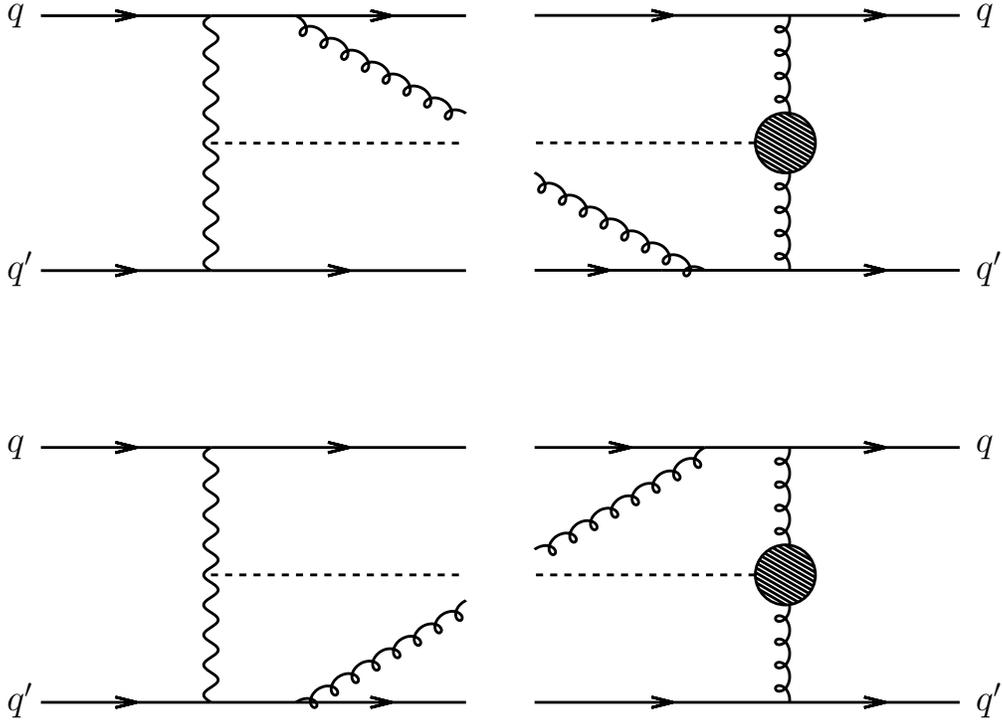,width=0.8\textwidth,clip=}
} 
\caption
{\label{fig:vbf-gf-re} 
Representative cut amplitudes for the $qq'\to qq'gH$ process via the
interference of WBF and GF amplitudes, at 
order $\mc{O}(\alpha^2 \alpha_s^3)$.
}
\end{figure}

\begin{figure}[!tb] 
\centerline{ 
\epsfig{figure=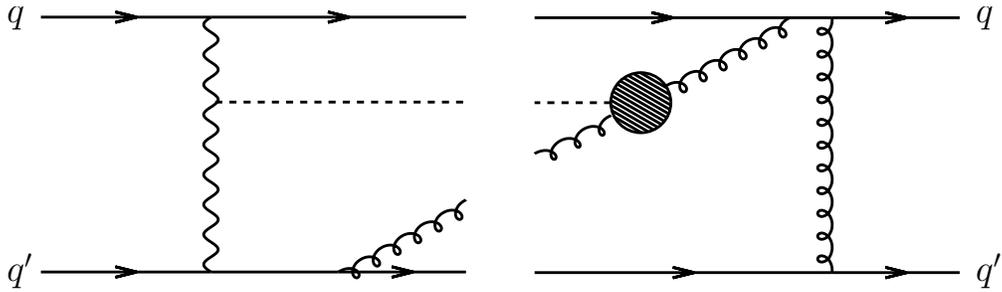,width=0.8\textwidth,clip=}
} 
\caption
{\label{fig:gf-ggh} 
Representative diagram contributing to the interference of 
$qq'\to qq'gH$ via WBF and GF, respectively, at 
order $\mc{O}(\alpha^2 \alpha_s^3)$.
}
\end{figure}

At the same order in the perturbative expansion, 
real emission diagrams have to be considered. Non-vanishing neutral-current 
contributions to the $qq'\to qq'gH$ process arise
from the interference of scattering diagrams like those depicted in 
Fig.~\ref{fig:vbf-gf-re}. 
%
% 
%%%%%%%%%%%%%%%%%%%%%%%%%%%%%%%%%%%%%%%%%%%%%%%%%%%%%%%%%%%%%%%%%%%%%%
%
% 
%%%%%%%%%%%%%%%%%%%%%%%%%%%%%%%%%%%%%%%%%%%%%%%%%%%%%%%%%%%%%%%%%%%%%%
%
The upper diagram shows an interference between gluon emission from the
$q$-line in the WBF amplitude and from the $q'$-line in the GF amplitude. In the
lower diagram the inverse configuration is illustrated. 
Interference graphs where both gluons are emitted from the same
quark  line cancel out when colors are summed over.
The same applies to those graphs of the GF amplitude where a gluon is attached
to the internal gluon line or to the $Hgg$ vertex. 
We denote the real emission amplitudes for $qq'\to qq'gH$ that do not cancel by 
$\mc{M}_{\rm{WBF}}^{(\rm{real})}$ and
$\mc{M}_{\rm{GF}}^{(\rm{real,t})}$, respectively. 
Further contributions to $qq'\to qq'gH$ scattering via GF, referred to as 
$\mc{M}_{\rm{GF}}^{(\rm{real,f})}$,  
arise from a topology absent in $qq'\to qq'H$, where the Higgs
boson is radiated off the final-state gluon rather than the $t$-channel exchange
boson, see Fig.~\ref{fig:gf-ggh}. 
%
% 
%%%%%%%%%%%%%%%%%%%%%%%%%%%%%%%%%%%%%%%%%%%%%%%%%%%%%%%%%%%%%%%%%%%%%%
%
% 
%%%%%%%%%%%%%%%%%%%%%%%%%%%%%%%%%%%%%%%%%%%%%%%%%%%%%%%%%%%%%%%%%%%%%%
%

Since only diagrams with gluons being emitted from different quark lines
contribute to the real emission, no collinearly divergent configurations
emerge. Singularities arise, however, when the final-state gluon in 
$\mc{M}_{\rm{WBF}}^{(\rm{real})}$ or $\mc{M}_{\rm{GF}}^{(\rm{real,t})}$ is soft.
Such divergences in the real emission diagrams are eventually 
canceled by respective singularities in the virtual contributions. To isolate
them in intermediate steps of the calculation, a proper regularization
scheme has to be utilized. We therefore perform our calculation in $d=4-2\eps$
dimensions, implementing both the dimensional regularization and the dimensional
reduction prescriptions~\cite{DRED}. Checking that both schemes yield the same
results provides a test of our calculation.

%%%%%%%%%%%%%%%%%%%%%%%%
%
\subsection{Virtual contributions}
For the discussion of the loop contributions to $Hjj$ production via WBF and GF,
respectively, we resort to the quark-quark scattering process
\beq
\label{eq:qqqqh}
q(p_a) + q'(p_b) \to q(p_1) + q'(p_2) + H(p_H)\,.
\eeq
The one-loop amplitudes we are considering, 
\bea
\mc{M}_{\rm{WBF}}^{(\rm{1-loop})} &=& 
\sum_{k=1}^{13} F_k^{\rm{WBF}}(p_a,p_1;p_b,p_2)\hat{\mc{M}}_{k}\:,
\\
\mc{M}_{\rm{GF}}^{(\rm{1-loop})} &=&
\sum_{k=1}^{13} F_k^{\rm{GF}}(p_a,p_1;p_b,p_2)\hat{\mc{M}}_{k}\;, 
\eea
can be expressed as linear combinations of process-dependent 
pre-factors, $F_k^{\rm{WBF}}$ and $F_k^{\rm{GF}}$, 
and fermion spinor chains  $\hat{\mc{M}}_k$, so-called 
``standard-matrix elements'' (SME). Following Ref.~\cite{ddrw:vvh}, 
we introduce
\bea
\Gamma^{qq}_{\{\al,\al\be\ga\}} &=&
\bar u(p_1,\lambda_1)\left\{\ga_\al,\ga_\al\ga_\be\ga_\ga\right\}
u(p_a,\lambda_a)\:,
\\
\Gamma^{q'q'}_{\{\al,\al\be\ga\}} &=&
\bar u(p_2,\lambda_2)\left\{\ga_\al,\ga_\al\ga_\be\ga_\ga\right\}
u(p_b,\lambda_b)\;,
\eea
where $u(p_i,\lambda_i)$ denotes the quark spinor for
fermion $i$ with momentum $p_i$ and helicity $\lam_i=\pm 1/2$. 
For contractions with an arbitrary  momentum $p$ we use the 
shorthand notation 
$\Gamma_{p}\equiv \Gamma_\mu p^\mu$. 
For the reaction~(\ref{eq:qqqqh}), 13~SME emerge, 
\beq
\def\arraystretch{1.5}
\begin{array}[b]{rclcrcl}
\hat{\mc{M}}_{\{1,2\}} &=&
\Ga^{qq}_{\al} \; \Ga^{q'q',\{\al,\al p_1 {p}_a\}}, 
& \qquad &
\hat{\mc{M}}_{\{3,4\}} &=& 
\Ga^{qq}_{\al p_2 p_b} \; \Ga^{q'q',\{\al,\al p_1 {p}_a\}}, 
\\
\hat{\mc{M}}_{\{5,6\}} &=& 
\Ga^{qq}_{p_b} \; \Ga^{q'q',\{{p}_a,p_1\}}, 
& \qquad &
\hat{\mc{M}}_{\{7,8\}} &=& 
\Ga^{qq}_{p_2} \; \Ga^{q'q',\{{p}_a,p_1\}}, 
\\
\hat{\mc{M}}_{\{9,10\}} &=& 
\Ga^{qq}_{\al\be p_b} \; \Ga^{q'q',\{\al\be {p}_a,\al\be p_1\}}, 
& \qquad &
\hat{\mc{M}}_{\{11,12\}} &=& 
\Ga^{qq}_{\al\be p_2} \; \Ga^{q'q',\{\al\be {p}_a,\al\be p_1\}}, 
\\
\hat{\mc{M}}_{13} &=& 
\Ga^{qq}_{\al\be\ga} \; \Ga^{q'q',\al\be\ga}. 
&&&&
\end{array}
\eeq
The SME are computed in two independent ways by means of the helicity
amplitude formalism of Ref.~\cite{HZ} and the Weyl--van der Waerden formalism of 
Ref.~\cite{Dittmaier:1998nn}, respectively.

The coefficients $F_k^{\rm{WBF}}$ and $F_k^{\rm{GF}}$ contain
coupling factors and remnants of scalar and tensor loop integrals,  
up to rank two and up to five propagator
denominators. 
For the computation of the tensor coefficients we have employed 
two different methods and developed two completely independent computer codes. 
These codes agree with a relative accuracy better than $10^{-8}$ for
non-exceptional phase-space points away from the zeroes of the Gram determinant. 
The basic
features of the two implementations are described in the following.

%
%
%%%%%%%%
%
\noindent
{\underline{\em Passarino--Veltman type tensor reduction}}:\\[2mm]
In one of our implementations, we have used the conventional 
Passarino--Veltman (PV) reduction formalism
\cite{PV,been:phd} for the computation of tensor integrals up to boxes, 
generalizing 
the method in a straightforward manner to pentagons in the framework of
dimensional regularization. All tensor
coefficients are expressed in terms of scalar master integrals in 
$(4-2\eps)$ dimensions with a regularization scale $\mu$. 
Singularities are manifested as single or double  poles in $\eps$.  
For GF, only integrals with vanishing internal masses emerge. Expressions
for the respective four-point integrals with up to two external 
off-shell legs are taken from
Ref.~\cite{dn:box}. The infrared divergent two- and three-point 
integrals can be
extracted from Refs.~\cite{been:phd,ditt:sing}.  
For WBF, scalar integrals with up to two massive propagators are needed. 
We have calculated the divergent box integrals by extracting
corresponding expressions in mass regularization from 
Refs.~\cite{Denner:1991qq,Beenakker:1988jr}.
Employing the method described in Ref.~\cite{ditt:sing} these expressions are
then transformed to dimensional regularization (see App.~\ref{app:int}) 
\footnote{We identified a misprint in one of the respective expressions of
Ref.~\cite{abhs:int}.}.
The remaining finite scalar integrals were calculated with 
{\tt LoopTools}~\cite{looptools}, and have been compared numerically to the
expressions given in Ref.~\cite{abhs:int}. 

The double and single pole terms of the coefficients $F_k^{\rm{WBF}}$ and 
$F_k^{\rm{GF}}$ are calculated analytically with the help of {\tt
mathematica}. To this end, 
we perform the tensor reduction in two steps: first for the singular pieces in
$d=4-2\eps$ dimensions, and then in four dimensions for the non-singular terms. 
For tensor integrals up to rank two this separation is trivial, since
poles do not mix with the finite terms 
when the reduction formalism is applied.  
Due to the absence of collinear
configurations in the interference process we are focusing on, all double poles
present in divergent  scalar integrals
cancel in the full loop amplitudes. This cancelation occurs for 
each coefficient $F_k^{\rm{GF}}, F_k^{\rm{WBF}} (k=1,\ldots, 13)$ separately. 
The remaining single poles for the GF and WBF contributions are  proportional to the 
respective tree-level amplitudes such that the singular parts of the loop-induced 
interference contribution take the form 
\bea
\label{eq:virt-sing}
 \overline{\sum}
 &&
 \hs{-4ex}
 2\mr{Re}\left[\mc{M}_{\rm{WBF}}^{(\rm{1-loop})}
			    \mc{M}_{\rm{GF}}^{(0)\star}
              +\mc{M}_{\rm{GF}}^{(\rm{1-loop})}
			    \mc{M}_{\rm{WBF}}^{(0)\star}\right]_\mr{sing}
\non\\			    
&=&
-\frac{\alpha_s}{2\pi} \,\Gamma(1+\eps)\,\mu^{2\eps}\,
\,\frac{1}{\eps}\,\,
 \overline{\sum}\,
 2\mr{Re}\left[\mc{M}_{\rm{WBF}}^{(0)}
			    \mc{M}_{\rm{GF}}^{(0)\star}	
	       +\mc{M}_{\rm{GF}}^{(0)}
			    \mc{M}_{\rm{WBF}}^{(0)\star}\right]
\non\\
&&\times
\left[
\ln\left(\frac{ s_{ab}}{\fpi\mu^2}\right) - 
\ln\left(\frac{-s_{a2}}{\fpi\mu^2}\right) - 
\ln\left(\frac{-s_{b1}}{\fpi\mu^2}\right) + 
\ln\left(\frac{ s_{12}}{\fpi\mu^2}\right)
\right]\,,
%\non
%\\
\eea
where $\overline{\sum}$ denotes averaging over initial-state spin degrees of
freedom and summation over final-state ones. 
Here, we have introduced the notation $s_{ij}=2p_i\cdot p_j$. 
We will see below that the divergent pieces are
canceled exactly by respective poles in the real emission contributions. 
In our Monte-Carlo program we thus retain only the finite parts of the 
scalar integrals. The tensor reduction for the non-singular terms can
then be performed numerically.

\noindent
{\underline{\em Denner--Dittmaier type tensor reduction}}:\\
In an independent implementation, we have performed the reduction from 
five-point to four-point integrals by means of the Denner--Dittmaier (DD)
reduction formalism \cite{DD:red}, while still reducing the four-point
integrals with the conventional PV tensor reduction discussed
above. 
Like the PV method the DD reduction is formulated in $d$ 
dimensions, but has the advantage of avoiding subtle cancelations, which can
spoil the numerical integration. 
Such cancelations originate from  
terms with a very small Gram determinant in the denominator. 
If two or more external momenta are linearly dependent, the Gram
determinant vanishes and we encounter numerical instabilities in the respective
phase-space regions.
In the PV approach, one power of the Gram determinant appears, in
general, in each step of the iterative reduction to lower-rank tensors, 
and also when an $N$-point function is reduced to $(N-1)$-point integrals.
In contrast to the PV reduction, the DD approach avoids Gram
determinants in the denominator in the reduction from five-point to four-point
integrals. 
In addition, it reduces the rank of the emerging four-point tensor integrals by
one.  A similar reduction formalism has also been given in
\cite{Binoth:2005ff}. 

As expected, the DD method turns out to be numerically far more stable
than the PV reduction.
Throughout our numerical studies, we will therefore resort to the DD
formalism. The PV reduction is used only to test our results. 

%%%%%%%%%%%%%%%%%%%%%%%%
%
\subsection{Real emission  contributions}
For the real emission contribution the partonic
subprocess
\beq
\label{eq:qqqqgh}
q(p_a) + q'(p_b) \to q(p_1) + q'(p_2) + g(p_g) + H(p_H)
\eeq
has  to be considered. 
The amplitudes 
$\mc{M}_{\rm{WBF}}^{(\rm{real})}$ and 
$\mc{M}_{\rm{GF}}^{(\rm{real})}=
 \mc{M}_{\rm{GF}}^{(\rm{real,t})}+\mc{M}_{\rm{GF}}^{(\rm{real,f})}$
are computed numerically by means of the helicity
amplitude formalism of Ref.~\cite{HZ}. Results for 
$q\bar q'\to q\bar q' gH$, $\bar q q'\to \bar q q' gH$, and $\bar q\bar q'\to \bar
q\bar q' gH$ are obtained analogously. Due to color conservation, 
crossing-related subprocesses with a gluon in the initial state such as 
$g q'\to q\bar q q' H$ do 
not contribute to the interference cross section.  

To test our implementation we have
compared our results for  $\mc{M}_{\rm{WBF}}^{(\rm{real})}$ and
$\mc{M}_{\rm{GF}}^{(\rm{real})}$ 
at the amplitude level and for the interference contribution   
$\overline{\sum} 
2\mr{Re}\left[\mc{M}_{\rm{WBF}}^{(\rm{real})}
\mc{M}_{\rm{GF}}^{(\rm{real})\star}+\mc{M}_{\rm{GF}}^{(\rm{real})}
\mc{M}_{\rm{WBF}}^{(\rm{real})\star}\right]$ 
to {\tt madgraph}~\cite{madgraph} and found complete agreement 
within the numerical accuracy of our program.

%%%%%%%%%%%%%%%%%%%%%%%%
%
\subsection{Subtraction procedure}
The real emission contributions contain soft divergences 
which  eventually cancel the corresponding poles in the virtual
contributions, cf.~Eq.~(\ref{eq:virt-sing}).
A convenient method for isolating the singularities is the so-called
phase-space slicing procedure. It relies on splitting the $qq'gH$ phase space 
into soft and hard regions by a suitable cutoff parameter and performing the
integration of the real emission contributions in the two regimes separately. 
To check our results we implement two conceptually different
slicing methods: The two-cutoff slicing method of Ref.~\cite{two-cutoff-pss} 
which has been developed in the context of mass regularization, and the 
phase-space slicing method of Ref.~\cite{one-cutoff-pss} which utilizes a 
Lorentz-invariant cutoff and dimensional regularization. 

\subsubsection{Lorentz-invariant phase-space slicing}
The slicing method of Ref.~\cite{one-cutoff-pss} divides
the phase space of the final-state particles into a hard region where all 
partons can be resolved and an infrared region for soft and collinear
configurations. In general, special care is necessary to separate soft and collinear regions
in order to avoid double counting of singular configurations. Since the
interference contributions of our interest are free of collinear singularities, the
formalism can be greatly simplified, however. In the case of $qq'\to qq'gH$ the
gluon is considered as infrared when 
\beq
s_{ig} = 2 p_i\cdot p_g < s_\mr{min}\,, \quad\mr{with}\;
i=a,b,1,2
\eeq
for an arbitrarily small cutoff parameter $s_\mr{min}$, where we closely follow
the notation of Ref.~\cite{tth:reina}. While for collinear
configurations only one $s_{ig}$ is small, the soft region is defined by
requiring at least two invariants to be smaller than $s_\mr{min}$. 
The partonic real emission cross
section can then be decomposed into a soft and a hard part, 
\beq
\hat{\sigma}^\mr{real} = \hat{\sigma}^\mr{soft} + \hat{\sigma}^\mr{hard} \,.
\eeq
The integration over the gluonic degrees of freedom is performed analytically in
$\hat{\sigma}^\mr{soft}$, but purely numerically in $\hat{\sigma}^\mr{hard}$. 

In order to calculate $\hat{\sigma}^\mr{soft}$ we use the factorization properties of
the real emission amplitude in the soft limit. As the energy of the emitted
gluon becomes small, the $qq'\to qq'gH$ interference amplitudes can be 
approximated by the tree-level interference amplitudes multiplied by a sum 
of eikonal terms,
\bea
\label{eq:soft-me}
\overline{\sum}
2\mr{Re}
\left[
\mc{M}_{\rm{WBF}}^{(\mr{real})} 
\mc{M}_{\rm{GF}}^{(\mr{real})\star} 
\right.
&&
\left.
\hs{-6ex}
+\mc{M}_{\rm{GF}}^{(\mr{real})} 
\mc{M}_{\rm{WBF}}^{(\mr{real})\star} 
\right]_\mr{soft}
\non\\
&=&
(4\pi\alpha_s)\,\mu^{2\eps}\,
\overline{\sum}\,
2\mr{Re}
\left[
\mc{M}_{\rm{WBF}}^{(0)} 
\mc{M}_{\rm{GF}}^{(0)\star} 
+\mc{M}_{\rm{GF}}^{(0)} 
\mc{M}_{\rm{WBF}}^{(0)\star} 
\right]
\non\\
&&
\times
\left[
 \frac{2s_{ab}}{s_{ag} s_{bg}}
-\frac{2s_{a2}}{s_{ag} s_{2g}}
-\frac{2s_{b1}}{s_{bg} s_{1g}}
+\frac{2s_{12}}{s_{1g} s_{2g}}
\right]\,.
\eea
The color structure of the soft contribution will be considered below.  
In the soft region, the four-particle $qq'gH$ phase space factorizes
into a three-particle $qq'H$ phase space and the soft gluon phase space for the
respective configuration, 
\beq
\left[d(PS_4)\right]^\mr{soft} = d(PS_3)\,d(PS_g)^\mr{soft}(i,j,g)\,,  
\eeq
where $d(PS_3)$ contains the flux factor, $1/(2\hat{s})$,  
with $\hat{s}$ denoting the partonic center-of-mass energy squared. 
For two outgoing partons $i$ and $j$, $d(PS_g)^\mr{soft}(i,j,g)$ is 
given by~\cite{one-cutoff-pss}
\bea
d(PS_g)^\mr{soft}(i,j,g) &=&
\frac{(4\pi)^\eps}{16\pi^2}\frac{s_{ab}^{2\eps-1}}{\Gamma(1-\eps)}
\left[s_{ig}s_{jg}s_{ij}\right]^{-\eps}
ds_{ig} ds_{jg}
\theta(s_\mr{min}-s_{ig})\,\theta(s_\mr{min}-s_{jg})\,.
\non\\
\eea
The integration over the soft gluon phase space can be performed for each term
in the soft interference amplitude Eq.~(\ref{eq:soft-me}) explicitely, using 
\bea
g_s^2 \mu^{2\eps} 
\int d(PS_g)^\mr{soft}(i,j,g)\frac{2s_{ij}}{s_{ig} s_{jg}} &=&
\frac{g_s^2}{8\pi^2}\frac{1}{\Gamma(1-\eps)}
\left(\frac{4\pi\mu^2}{s_\mr{min}}\right)^\eps
\frac{1}{\eps^2}
\left(\frac{s_{ij}}{s_\mr{min}}\right)^\eps\,.
\eea
The generalization of this expression to cases where one of the partons 
$i,j$ is incoming rather than outgoing is straightforward. The soft part 
of the real emission cross section  then takes the form 
\bea
\label{eq:sigma-soft}
\hat{\sigma}^\mr{soft} &=& 
\frac{\ca\cf}{2}
\int \left[d(PS_4)\right]^\mr{soft}
\overline{\sum}
2\mr{Re}
\left[
\mc{M}_{\rm{WBF}}^{(\mr{real})} 
\mc{M}_{\rm{GF}}^{(\mr{real})\star} 
+
\mc{M}_{\rm{GF}}^{(\mr{real})} 
\mc{M}_{\rm{WBF}}^{(\mr{real})\star} 
\right]_\mr{soft}
\non\\ 
&=& 
\frac{\alpha_s}{2\pi}\, \Gamma(1+\eps)\,
\frac{\ca\cf}{2}
\int d(PS_3)\,
\overline{\sum}
2\mr{Re}
\left[
\mc{M}_{\rm{WBF}}^{(0)} 
\mc{M}_{\rm{GF}}^{(0)\star} 
+
\mc{M}_{\rm{GF}}^{(0)} 
\mc{M}_{\rm{WBF}}^{(0)\star} 
\right]
\non\\
&&\times
\Biggl\{
\left(
\frac{1}{\eps}
+\ln \frac{4\pi\mu^2}{s_\mr{min}}
%+\ln\left(\frac{4\pi\mu^2}{s_\mr{min}}\right)
\right)
\cdot
\left[
 \ln\left( \frac{s_{ab}}{s_\mr{min}}\right)
-\ln\left(-\frac{s_{a2}}{s_\mr{min}}\right)
-\ln\left(-\frac{s_{b1}}{s_\mr{min}}\right)
+\ln\left( \frac{s_{12}}{s_\mr{min}}\right)
\right]
\non\\
&&
\hs{.4cm}
+\frac{1}{2}
\left[
 \ln^2\left( \frac{s_{ab}}{s_\mr{min}}\right)
-\ln^2\left(-\frac{s_{a2}}{s_\mr{min}}\right)
-\ln^2\left(-\frac{s_{b1}}{s_\mr{min}}\right)
+\ln^2\left( \frac{s_{12}}{s_\mr{min}}\right)
\right]
\Biggr\}\,,
\non\\
\eea
where we have included the color factor $\ca\cf/2=2$. 
When $\hat{\sigma}^\mr{soft}$ is combined with the virtual contributions,
\beq
\hat{\sigma}^\mr{qq'H} =\hat{\sigma}^\mr{virt} + \hat{\sigma}^\mr{soft}\,,
\eeq
all $1/\eps$ poles cancel [cf.~Eq.~(\ref{eq:virt-sing})]. 
The remaining terms are
finite and can be integrated over the three-particle phase space of the $qq'H$ 
system and convoluted with the parton distributions of the incoming fermions 
numerically. The resulting three-particle contribution, $\sigma^\mr{qq'H}$,  
depends on the unphysical cutoff parameter $s_\mr{min}$. This dependence cancels,
however, once $\sigma^\mr{qq'H}$ is combined with the hard part of the real
emission cross section, 
\beq
\sigma^\mr{full} =\sigma^\mr{hard} + \sigma^\mr{qq'H}\,.
\eeq
Checking that the full NLO interference cross section is independent of the
cutoff parameter therefore provides another  important test of our calculation.

%%%%%%%%%%%%%%%%%%%%%%%%%%%%%%%%%%%%%%%%%%%%%%%%%%%
%
\subsubsection{Phase-space slicing with energy-cutoff}
The phase-space slicing method of Ref.~\cite{two-cutoff-pss} in principle 
requires two cutoff parameters for separating finite from collinear and 
soft divergent regions. For our application, however, no collinear 
singular configurations emerge. Thus, applying a single cutoff on the 
energy of the potentially soft gluon is sufficient.    

In analogy to the Lorentz-invariant slicing method described previously, 
the real emission contribution can be evaluated numerically in the ``hard'' 
region of phase space above the cutoff, where it is completely finite. 
Below the energy cutoff, however, the phase-space integration over the 
gluonic degrees of freedom is performed analytically.
Following Refs.~\cite{racoonww,denner:habil}, this ``soft'' contribution to 
the partonic cross section can be written as 
\bea
\label{eq:2cutoffgen}
\hat{\sigma}^{\rm soft,E-cut}& = & 
\frac{\alpha_s}{2\pi}\,
\frac{\ca\cf}{2}
\int d(PS_3)\,
\overline{\sum}
2\mr{Re}
\left[
 \mc{M}_{\rm{WBF}}^{(0)} 
 \mc{M}_{\rm{GF}}^{(0)\star} 
+
 \mc{M}_{\rm{GF}}^{(0)} 
 \mc{M}_{\rm{WBF}}^{(0)\star} 
\right]
\non\\
&&\times
\int_{E_3<\Delta E \atop {|{\bf p_g}|^2}=E_3^2-\lambda^2} 
           \frac{d^3{\bf p_g}}{2\pi E_3}
\left[
 \frac{2s_{ab}}{s_{ag} s_{bg}}
-\frac{2s_{a2}}{s_{ag} s_{2g}}
-\frac{2s_{b1}}{s_{bg} s_{1g}}
+\frac{2s_{12}}{s_{1g} s_{2g}}
\right]\,,
\eea
where $E_3$ is the gluon energy, $\Delta E$ the energy cutoff in the 
rest frame of the two incoming partons and $\lambda$ a mass used as regulator.
As above, $d(PS_3)$ denotes the $qq'H$ phase space. 
Evaluating the integrals over the gluonic degrees of 
freedom in Eq.~(\ref{eq:2cutoffgen}) and rewriting the mass-regulated result 
in terms of dimensional regularization, the corresponding soft cross section 
is of the form   
\bea
\label{eq:2cutoff}
\hat{\sigma}^{\rm soft,E-cut} &=&
\frac{\alpha_s}{2\pi} \,
\frac{\ca\cf}{2}
\int d(PS_3)\,
\overline{\sum}
2\mr{Re}
\left[
 \mc{M}_{\rm{WBF}}^{(0)} 
 \mc{M}_{\rm{GF}}^{(0)\star} 
+
 \mc{M}_{\rm{GF}}^{(0)} 
 \mc{M}_{\rm{WBF}}^{(0)\star} 
\right]
\non\\
&&\hs{-4ex}
\times
\left\{
  \frac{\Gamma(1+\eps)}{\eps}\left(\frac{\pi\mu^2}{\Delta E^2}\right)^{\eps}
\right.
%\non\\
%&&\times
\left[
\ln\left(\frac{ s_{ab}}{\fpi\mu^2}\right) - 
\ln\left(\frac{-s_{a2}}{\fpi\mu^2}\right) - 
\ln\left(\frac{-s_{b1}}{\fpi\mu^2}\right) + 
\ln\left(\frac{ s_{12}}{\fpi\mu^2}\right)
\right]\,
\non\\
&& 
%\hs{-4ex}
%\left.
   + {\rm Li_2}\left(1-\frac{4 E_a E_b}{s_{ab}}\right)
   - {\rm Li_2}\left(1-\frac{4 E_a E_2}{s_{a2}}\right) 
\non\\
&& 
\left.
   - {\rm Li_2}\left(1-\frac{4 E_b E_1}{s_{b1}}\right) 
   + {\rm Li_2}\left(1-\frac{4 E_1 E_2}{s_{12}}\right) 
\right\},
\non\\
\eea
where the $E_i$ denote the quark energies in the partonic rest frame.
In complete analogy to the Lorentz-invariant phase-space slicing, 
the soft contribution to the partonic cross section is combined with 
the virtual cross section. The resulting sum is then free of soft poles 
and can be evaluated numerically. Upon adding the hard part of the real 
emission contribution, the  dependence on the cutoff parameter $\Delta E$
cancels. 

%%%%%%%%%%%%%%%
%
\subsubsection{Checks}
We have checked that the total $pp\to Hjj$ interference cross section
at the LHC within typical WBF cuts 
(for details, see our standard definition of cuts in
Sec.~\ref{sec:res}) is independent of the cutoff parameter for both phase-space 
slicing schemes. 

\begin{figure}[!tb] 
\centerline{ 
\epsfig{figure=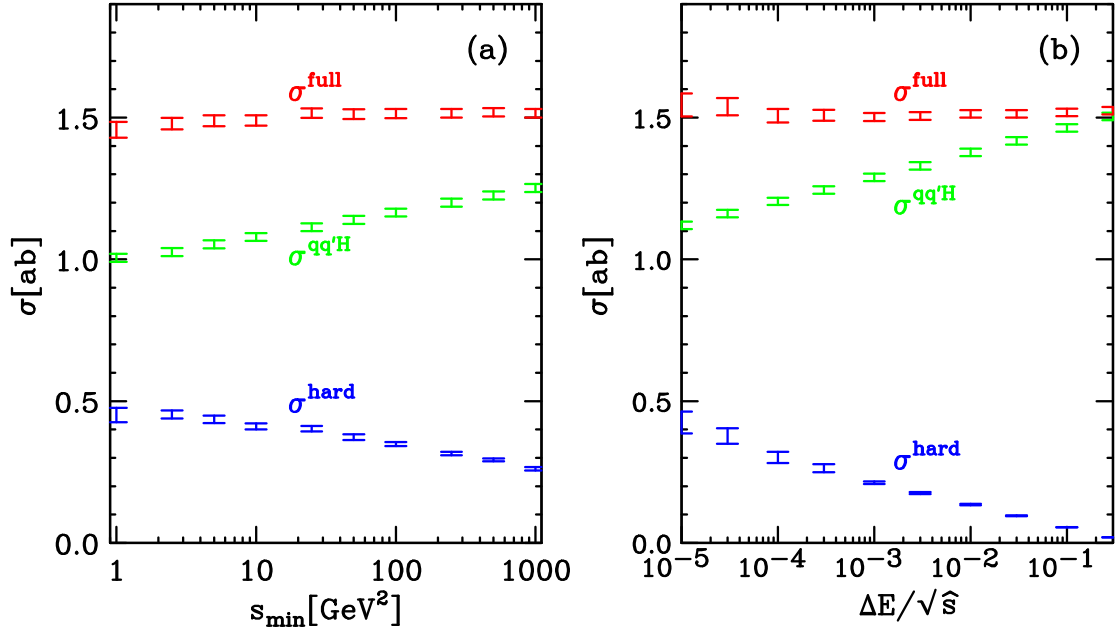,width=0.9\textwidth,clip=}
} 
\caption
{\label{fig:smin-slice} 
Dependence of the interference cross section for $pp\to Hjj$ production at the
LHC within standard %WBF 
selection cuts on the cutoff of the
Lorentz-invariant phase-space slicing method (a) and of the energy-cutoff slicing method (b). Shown are $\sigma^\mr{hard}$
(blue), $\sigma^\mr{qq'H}$ (green), and their
sum, $\sigma^\mr{full}$ (red). 
}
\end{figure}

For the Lorentz-invariant slicing method, we have varied $s_\mr{min}$ 
in the range 
$1$~GeV$^2<s_\mr{min}<10^{3}$~GeV$^2$. For smaller cutoff values, large 
logarithms arise and numerical instabilities are
to be expected. If, on the other hand, a very large value is chosen for 
$s_\mr{min}$, the soft approximation used for determining $\hat{\sigma}^\mr{soft}$ is
not applicable anymore. Fig.~\ref{fig:smin-slice}~(a) 
%
%%%%%%%%%%%%%%%%%%%%%%%%%%%%%%%%%%%%%%%%%%%%%%%%%%%%%%%%%%%%%%%%%%%%%%
%
% 
%%%%%%%%%%%%%%%%%%%%%%%%%%%%%%%%%%%%%%%%%%%%%%%%%%%%%%%%%%%%%%%%%%%%%%
%
demonstrates that the two
contributions $\sigma^\mr{qq'H}$ and  $\sigma^\mr{hard}$ individually depend on
$s_\mr{min}$, while the sum $\sigma^\mr{full}$ is constant in the considered range
of the cutoff parameter.  

A very similar pattern arises for the energy-cutoff slicing method, 
depicted in Fig.~\ref{fig:smin-slice}~(b). We have normalized the energy 
cutoff $\Delta E$ by  $\sqrt{\hat{s}}$ for this study.

%
%%%%%%%%%%%%%%%%%%%%%%%%%%%%%%%%%%%%%%%%%%%%%%%%%%%%%%%%%%%%%%%%%
\section{Numerical results}
\label{sec:res}

The cross-section contributions discussed above have been implemented in a
fully flexible parton level Monte-Carlo program, structured analogous to the
{\tt vbfnlo} code~\cite{vbfnlo} which has been developed for the study of WBF-type
production processes at the LHC. 

The loop-induced interference contributions for $Hjj$ production via GF and
WBF we consider are a gauge-invariant sub-class of the full NLO-QCD
corrections to the scattering process $pp\to Hjj$. For the parton distribution
functions of the proton we therefore use the CTEQ6M set at NLO \cite{cteq6} with
$\alpha_s(M_Z) = 0.118$. We set quark masses to zero throughout and neglect
contributions from external top or bottom quarks. 
%For the Cabbibo-Kobayashi-Maskawa matrix, $V_{CKM}$, we use the identity
%matrix which is equivalent to employing the exact $V_{CKM}$ when the summation
%over all quark flavors is performed, see Ref.~\cite{BJOZ:WZ}. 
%
As electroweak input parameters we have chosen $m_Z=91.188$~GeV,
$m_W=80.419$~GeV, and the measured value of $G_\mr{F} = 1.166\times
10^{-5}/$~GeV$^2$. 
Thereof, we compute $\sin^2\theta_\mr{W}$ and
$\alpha$ using LO electroweak relations. For reconstructing
jets from final-state partons, we use the $k_\mr{T}$ algorithm \cite{kT} with
resolution parameter $R^\mr{k_T}=0.8$. 

Since we want to study the impact of the interference contributions on the
Higgs signal in WBF, we apply cuts that are typical for WBF studies at the
LHC. We require at least two hard jets with
\beq
\label{eq:jcut}
p_{\mr{T}j}\geq 20~\mr{GeV}\,,\quad
|y_j|\leq 4.5\,,
\eeq
where $p_{\mr{T}j}$ is the transverse component and 
$y_j$ the rapidity of the (massive) jet momentum which is
reconstructed as the four-vector sum of massless partons of pseudorapidity
$|\eta|< 5$. We refer to the two reconstructed jets of highest transverse momentum as ``tagging jets''. 
The Higgs boson decay products, which we generically  call ``leptons'' in the
following, are required to be located between the two tagging jets and they
should be well observable. To simulate a generic Higgs decay without
specifying a particular channel we generate an
isotropic Higgs boson decay into two massless particles (which represent
$\gamma\gamma$ or $b\bar b$ final states) and require
\beq
\label{eq:lcut}
p_{\mr{T}\ell}\geq 20~\mr{GeV}\,,\quad
|\eta_\ell|\leq 2.5\,,\quad
\Delta R_{j\ell}\geq 0.6\,,
\eeq
where $\Delta R_{j\ell}$ denotes the jet-lepton separation in the
rapidity-azimuthal angle plane. In addition, the leptons need to fall between
the rapidities of the two tagging jets
\beq
y_{j,min}<\eta_\ell<y_{j,max}\,.
\eeq
Furthermore, we impose large rapidity separation of the two tagging jets,
\beq 
\Delta y_{jj}=|y_{j1}-y_{y2}|>4\,,
\eeq
and demand that the two tagging jets be located in opposite detector
hemispheres,
\beq
y_{j1}\times y_{j2}<0\,,
\eeq
with an invariant mass
\beq
\label{eq:jjcut}
M_{jj}>600~\mr{GeV}\,.
\eeq

To ensure the reliability of our calculation, we have compared our results to
those of  Ref.~\cite{abhs:int} and found agreement with their main predictions. 
Diagrams where the Higgs boson is radiated off the final-state gluon rather 
than  the $t$-channel exchange boson as in Fig.~\ref{fig:gf-ggh} have not
been considered in  \cite{abhs:int}. This approximation seems reasonable, as we
found that contributions from these graphs amount to only
about $0.3\%$ of the total interference cross section. For individual
subprocesses they can be larger, however. For the $dd\to ddH$ channel, for
instance, they yield approximately $5\%$ of the subprocess-cross section. 

\begin{figure}[!tb] 
\centerline{ 
\epsfig{figure=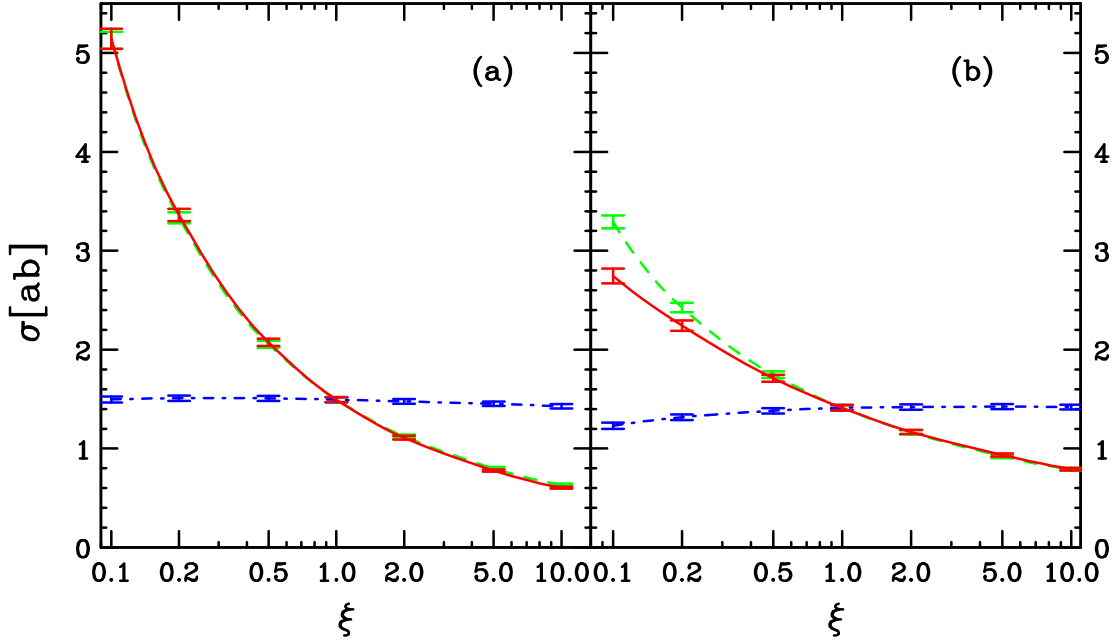,width=0.9\textwidth,clip=}
} 
\caption
{\label{fig:scale} 
Dependence of the total interference cross section $\sigma_\mr{int}^\mr{cuts}$ 
for $Hjj$ production at the LHC on the factorization and renormalization 
scales for the two different scenarios described in the text. 
The factorization scale $\mu_\mr{f}$ and the renormalization scale $\mu_\mr{r}$
are scaled as $m_H$ for (a) and as the jets' transverse momenta in (b), cf.\
Eqs.~(\ref{eq:scale_mh}) and (\ref{eq:scale_pt}), respectively. 
The curves show
$\sigma_\mr{int}^\mr{cuts}$ as a function of the scale parameter $\xi$ for three
different cases: $\xi_\mr{r}=\xi_\mr{f}=\xi$ (solid red),  $\xi_\mr{f}=\xi$ and
$\xi_\mr{r}=1.0$ (dot-dashed blue), $\xi_\mr{r}=\xi$ and $\xi_\mr{f}=1.0$ (dashed
green). 
}
\end{figure}

In Fig.~\ref{fig:scale}
%
% 
%%%%%%%%%%%%%%%%%%%%%%%%%%%%%%%%%%%%%%%%%%%%%%%%%%%%%%%%%%%%%%%%%%%%%%
%
% 
%%%%%%%%%%%%%%%%%%%%%%%%%%%%%%%%%%%%%%%%%%%%%%%%%%%%%%%%%%%%%%%%%%%%%%
%
%
we show the total cross section $\sigma_\mr{int}^\mr{cuts}$ 
for the interference contribution within the
cuts of Eqs.~(\ref{eq:jcut})--(\ref{eq:jjcut}) and for a Higgs mass of
$m_H=120$~GeV. The factorization scale, $\mu_\mr{f}$, and the renormalization scale,
$\mu_\mr{r}$,  which enters the strong coupling are chosen as follows: In panel~(a), we set 
\bea
\label{eq:scale_mh}
\mu_\mr{f} = \xi_\mr{f} m_H \,,\quad
\alpha_s^3(\mu_\mr{r}) =\alpha_s^3(\xi_\mr{r} m_H)\,.
\eea 
In panel~(b), we associate the scale for gluon emission from either quark line with the
transverse momentum of the corresponding jet by setting 
\bea 
\label{eq:scale_pt}
\mu_\mr{f} =\xi_\mr{f} p_{\mr{T}j}\,,\quad 
\alpha_s^3(\mu_\mr{r}) = 
\alpha_s(\xi_\mr{r} p_\mr{T1})\cdot\alpha_s(\xi_\mr{r} p_\mr{T2})\cdot \alpha_s(\xi_\mr{r} m_H)\,.
\eea  
Due to the absence of collinear singularities, $\mu_\mr{f}$ enters only via 
the parton distribution functions of the incoming fermions, which are mainly
probed at rather large values of Feynman~$x$. In this regime, 
the valence and sea quark distributions depend on the factorization scale 
only mildly. Thus, the variation of $\sigma_\mr{int}^\mr{cuts}$  
with $\mu_\mr{f}$ is very small. 
On the other hand, the interference cross section exhibits a pronounced
dependence on $\mu_\mr{r}$. Since the loop-induced 
GF$\times$WBF interference in  $qq'\to qq'H$ production,  
$\sigma_\mr{int}^\mr{cuts}$, represents the first
non-vanishing contribution in the perturbative expansion, the
renormalization scale enters only via the strong coupling constant. Thus, the
entire $\mu_\mr{r}$ dependence of the interference cross section can be traced back
to the variation of the $\alpha_s^3(\mu_\mr{r})$ coupling factor 
with the renormalization scale. 
Reminiscent of what has been observed for pure WBF production processes (cf.,
e.g., Ref.~\cite{BJOZ:WZ}) a dynamical scale choice as in 
Eq.~(\ref{eq:scale_pt}), see Fig.~\ref{fig:scale}(b), yields predictions
with a somewhat reduced scale dependence as compared to the fixed scale option
of Eq.~(\ref{eq:scale_mh}), shown in Fig.~\ref{fig:scale}(a).  

Compared to the total WBF cross section within typical selection cuts, the
interference contribution we have calculated is almost negligible in magnitude. 
%
%%%%%%%%%%%%%%%%%%%%%%%%%%%%%%%%
%
\renewcommand{\baselinestretch}{1.5}
\begin{table}[!tb]
\begin{center}
\begin{tabular}{|c|c|c|}
\hline
initial-state flavor combination
&$\sigma^\mr{cuts}_\mr{int}$[ab] & $\sigma^\mr{cuts}_\mr{WBF}$[fb]
\\
\hline
\hline
NC: $(u+c)(u+c)+(d+s)(d+s)$
&
51.4
&
72.3
\\
\hline
NC: $(u+c)(d+s)$
&
-49.8
&  
70.8 
\\
\hline
CC: $(u+c)(d+s)$
&
--
&
405.7
 \\
\hline
NC: $(u+c)(\bar u+\bar c)+(d+s)(\bar d+\bar s)$
&
-3.1
&
39.3
\\
\hline
NC: $(u+c)(\bar d+\bar s)+(\bar u+\bar c)(d+s)$
&
2.2
&
43.0
\\
 \hline
CC: $(u+c)(\bar u+\bar c)+(d+s)(\bar d+\bar s)$
&
--
&
230.7
 \\
\hline
NC: $(\bar u+\bar c)(\bar u+\bar c)+(\bar d+\bar s)(\bar d+\bar s)$
&
4.0
&
5.1
\\
\hline
NC: $(\bar u+\bar c)(\bar d+\bar s)$
&
-3.2
&
4.3
\\
\hline
CC: $(\bar u+\bar c)(\bar d+\bar s)$
&
--
&
25.7
\\
\hline
\hline
sum
&
1.5
&
896.9
\\
\hline
\end{tabular}
\end{center}
\renewcommand{\baselinestretch}{1.}
\caption{
\label{tab:xsec} 
Contributions of various neutral current (NC) flavor combinations to 
$\sigma^\mr{cuts}_\mr{int}$ (in ab) and $\sigma^\mr{cuts}_\mr{WBF}$ (in fb), 
and of the charged current (CC) contributions to WBF. Also shown is  their sum 
within the cuts of Eqs.~(\ref{eq:jcut})--(\ref{eq:jjcut}). 
}
\end{table}
\renewcommand{\baselinestretch}{1.}
%
%
%%%%%%%%%%%%%%%%%%%%%%%%%%%%%%%%
%
In Tab.~\ref{tab:xsec} 
we list $\sigma^\mr{cuts}$ for both, interference and pure
WBF cross sections for the various flavor combinations of the 
scattering quarks
and antiquarks, setting $\mu_\mr{f}=\mu_\mr{r}=m_H$. 
For WBF, we consider neutral and charged current subprocesses 
at $\mc{O}(\alpha^3)$. No $W$-exchange diagrams contribute to the 
interference cross section. Tab.~\ref{tab:xsec} reveals the   
strong cancelations occurring in $\sigma_\mr{int}^\mr{cuts}$ 
among the separate channels. While some contributions, in
particular for the $qq'$ subprocesses, are sizeable, their sum amounts to
1.5~ab only. We will show below that the subtle cancelation between the
same and opposite isospin $qq$, $q\bar q$, and $\bar q\bar q$ scattering
contributions leads to unexpected shapes of kinematic distributions in
flavor-blind experiments. 

\begin{figure}[!tb] 
\centerline{ 
\epsfig{figure=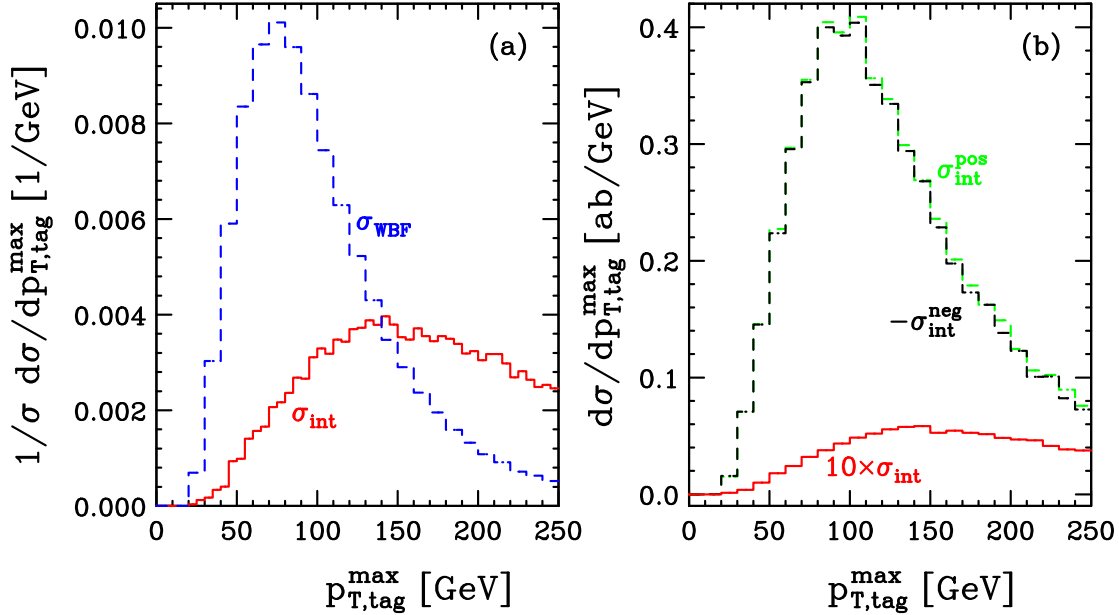,width=0.9\textwidth,clip=}
}
\caption
{\label{fig:ptmax}
Panel~(a) shows the 
normalized transverse momentum distributions 
for the tagging jet with the highest $p_\mr{T}$ for 
WBF (dashed blue line) and for the WBF$\times$GF 
interference contribution (solid red line).  
In panel~(b) the sum of all positive (dashed green line) and the magnitude of all negative contributions (dashed black line) are shown separately. The solid red line gives the sum of all contributions, multiplied by a factor of 10.
}
\end{figure}

Figure~\ref{fig:ptmax}
%
% 
%%%%%%%%%%%%%%%%%%%%%%%%%%%%%%%%%%%%%%%%%%%%%%%%%%%%%%%%%%%%%%%%%%%%%%
%
% 
%%%%%%%%%%%%%%%%%%%%%%%%%%%%%%%%%%%%%%%%%%%%%%%%%%%%%%%%%%%%%%%%%%%%%%
%
%
depicts  the shapes of the transverse-momentum 
distributions for ``pure'' WBF $Hjj$ production and 
for the  WBF$\times$GF interference contribution we have 
calculated. The very hard $p_\mr{T}$ distribution encountered for the 
interference  significantly differs from the shape of 
the WBF curve. The small size of the $p_\mr{T}$ distribution at low momentum transfer
is mainly due to strong cancelations among the different flavor contributions
to the full $pp\to Hjj$ interference cross section, as illustrated by
Fig.~\ref{fig:ptmax}(b), where the contributions 
for same isospin and opposite isospin 
$qq+q\bar q+\bar q\bar q$ scattering, 
$\sigma^\mr{pos}_\mr{int}$ and $\sigma^\mr{neg}_\mr{int}$, are shown separately.
The two contributions cancel almost
precisely to give the total interference contribution,
$\sigma^\mr{pos}_\mr{int}+\sigma^\mr{neg}_\mr{int}=\sigma_\mr{int}$. 
At high $p_\mr{T}$, the cancelation effects are less pronounced. 
In short, the interference contribution has a harder transverse-momentum spectrum
than expected, because of a very efficient cancelation around
$p_\mr{T,tag}^\mr{max}\sim 100$~GeV, where the individual distributions peak.

\begin{figure}[!tb] 
\centerline{ 
\epsfig{figure=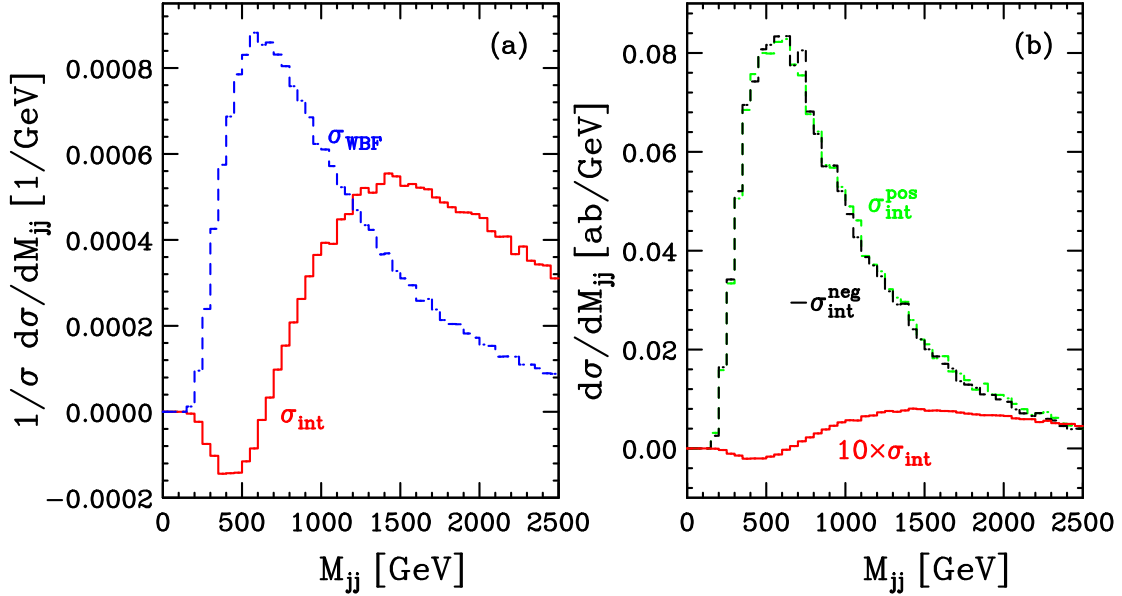,width=0.9\textwidth,clip=}
} 
\caption
{\label{fig:mjj} 
Panel~(a) shows the normalized tagging-jet invariant mass distributions 
for WBF (dashed blue line) and for the WBF$\times$GF 
interference contribution (solid red line).   
Panel~(b) depicts the sum of all positive contributions, $\sigma_\mr{int}^\mr{pos}$ (dashed green line),  
the magnitude of all negative contributions , $-\sigma_\mr{int}^\mr{neg}$ (dashed black line), and their sum $\sigma_\mr{int}$, multiplied by a factor of 10 (solid red line). 
}
\end{figure}

This cancelation pattern is reflected by the tagging-jet 
invariant mass $M_{jj}$. For studying the corresponding  
shapes of the pure EW and of the mixed QCD-EW production processes, 
we have switched off the invariant mass cut of 
Eq.~(\ref{eq:jjcut}). The emerging curves are 
displayed in Fig.~\ref{fig:mjj}.   
% 
%%%%%%%%%%%%%%%%%%%%%%%%%%%%%%%%%%%%%%%%%%%%%%%%%%%%%%%%%%%%%%%%%%%%%%
%
% 
%%%%%%%%%%%%%%%%%%%%%%%%%%%%%%%%%%%%%%%%%%%%%%%%%%%%%%%%%%%%%%%%%%%%%%
%
While the interference cross section is negative at small values of the
dijet-invariant mass, it is relatively large at high $M_{jj}$. Indeed, 
is is remarkable to find that the interference cross section yields an even
harder $M_{jj}$ distribution than the pure WBF cross section does. 
This behavior is somewhat unexpected if considering the rather soft
invariant mass distribution of the pure GF $Hjj$ production process which
has been reported in the literature \cite{GF:LO}. The full GF $pp\to Hjj$ cross
section, however, is dominated by gluon-initiated partonic channels such as
$gg\to ggH$ and $qg\to qgH$. To the interference cross section, on the other
hand, only  quark (and anti-quark) initiated subprocesses contribute, which tend
to give larger
values of $M_{jj}$ than gluonic contributions. More importantly, the cancelation
effects reported above in the context of the tagging-jet transverse-momentum distribution affect the summation over the 
various flavor contributions to the
dijet-invariant mass distribution in a similar manner, thereby giving rise to a
broad invariant-mass  distribution which is very small at low values of
$M_{jj}$. 

\begin{figure}[!tb] 
\centerline{ 
\epsfig{figure=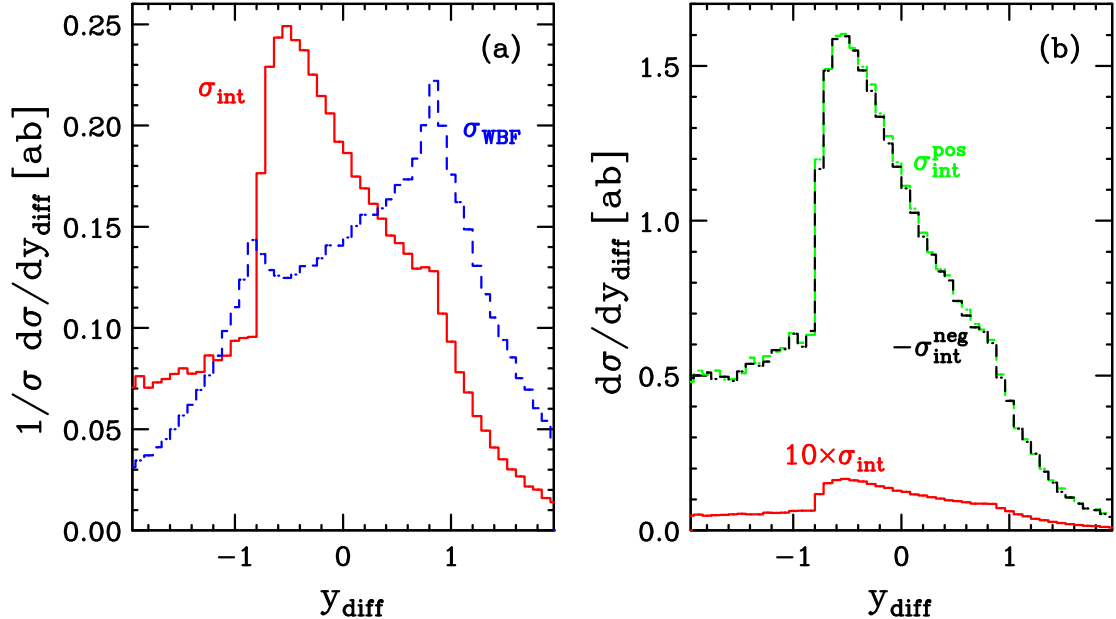,width=0.9\textwidth,clip=}
} 
\caption
{\label{fig:ydiff} 
Panel~(a) shows the 
normalized rapidity-separation distribution of the non-tagged jet for WBF
(dashed blue line) and for the  WBF$\times$GF 
interference contribution (solid red line).
Panel~(b) depicts the sum of all positive contributions, $\sigma_\mr{int}^\mr{pos}$ (dashed green line),  
the magnitude of all negative contributions, $-\sigma_\mr{int}^\mr{neg}$ (dashed black line), and their sum $\sigma_\mr{int}$, multiplied by a factor of 10 (solid red line). 
}
\end{figure}

The afore-mentioned cancelations have different effects on 
the rapidity distribution of the third, non-tagged jet with respect to
the tagged jet located in the positive-rapidity hemisphere,
\beq
y_\mr{diff} = y_3 - \mr{max}\,(y_1,y_2)\,,
\eeq 
which is shown in Fig.~\ref{fig:ydiff} 
% 
%%%%%%%%%%%%%%%%%%%%%%%%%%%%%%%%%%%%%%%%%%%%%%%%%%%%%%%%%%%%%%%%%%%%%%
%
% 
%%%%%%%%%%%%%%%%%%%%%%%%%%%%%%%%%%%%%%%%%%%%%%%%%%%%%%%%%%%%%%%%%%%%%%
%
for the interference contribution and the ``pure'' WBF cross section. 
The separation of the lowest $p_\mr{T}$-jet from the tagged jet located in
the negative-rapidity hemisphere, $-y_3 + \mr{min}\,(y_1,y_2)$,  
would be a mirror copy thereof due to our symmetric selection cuts. 
For generating the distribution, we required a minimum transverse momentum 
of $p_\mr{T3}\geq 10$~GeV for the third jet in addition to the 
selection cuts of Eqs.~(\ref{eq:jcut})--(\ref{eq:jjcut}). 
The peak of the distribution at small $|y_\mr{diff}|$ emphasizes that 
the ``soft'' jet prefers being close
in rapidity to the hard jet in the respective detector hemisphere for both,
interference and WBF contributions. While in WBF the third jet prefers
rapidities larger than the associated tag jet, $y_\mr{diff}>0$, 
for the interference contribution $y_\mr{diff}$ peaks at negative
values for the various flavor contributions and their sum, indicating that the soft jet is
typically located in {\em between} the two tagged jets.  
This may indicate that the rapidity gap for a color singlet EW-boson exchange
may in general be filled by the EW-QCD interference contribution. 
%

%
%%%%%%%%%%%%%%%%%%%%%%%%%%%%%%%%%%%%%%%%%%%%%%%%%%%%%%%%%%%%%%%%%
%

\section{Summary and conclusions}
\label{sec:concl}
In this article we have computed the order $\mc{O}(\alpha^2\alpha_s^3)$ 
interference contributions to the $Hjj$ production cross section in $pp$ 
collisions at the LHC via GF and WBF. Since results for the
total interference cross section and angular distributions 
have already been discussed in the literature \cite{abhs:int}, we have put special emphasis on
technical and phenomenological aspects of the calculation which have not been
discussed elsewhere. In particular, we have given a detailed outline of the
methods used for the evaluation  of loop contributions, the subtraction of
singularities present in intermediate steps of the calculation, and of the
checks we have performed to ensure the reliability of our results. In the real
emission contributions we have included a finite class of diagrams that has not been
considered previously. We found the numerical value of these contributions
small, however. 

Having implemented the interference amplitudes in a flexible Monte-Carlo program
based on the {\tt vbfnlo}  framework of Ref.~\cite{vbfnlo}, we are able to
provide total cross sections and arbitrary kinematic distributions within
experimentally feasible selection cuts. Considering the interference
cross section as possible ``contamination'' of the clean WBF $Hjj$ production
signature, we have studied the associated contributions within typical WBF cuts
with widely separated hard tag-jets and compared the shape of some
characteristic distributions to those of the respective pure WBF curves. We
found that, indeed, the interference contributions exhibit features rather
different from the WBF signal which are caused by strong cancelations among the
separate flavor channels. However, due to the small size of the
interference cross section which is found to be in the atto-barn range only, the
impact of this contribution to both, integrated cross sections and differential
distributions, is negligible. 

%%%%%%%%%%%%%%%%%%%%%%%%%%%%%%%%%%%%%%%%%%%%%%%%%%%%%%%%%%%%%%%%%
%%%%%%%%%%%%%%%%%%%%%%%%%%%%%%%%%%%%%%%%%%%%%%%%%%%%%%%%%%%%%%%%%
%
\section*{Acknowledgements}
We are grateful to Stefan Dittmaier and Dieter Zeppenfeld for helpful comments. 
A.\ B.\ would like to thank Junpei Fujimoto and Yoshimasa Kurihara for useful
discussions. 
Our work was supported by the Japan Society for the Promotion of
Science (JSPS), the Deutsche Forschungsgemeinschaft (DFG), and partly by the
Grant-in-Aid for scientific research (No. 17540281) from MEXT, Japan.
%
%%%%%%%%%%%%%%%%%%%%%%%%%%%%%%%%%%%%%%%%%%%%%%%%%%%%%%%%%%%%%%%%%
%
\newpage
\appendix
\section{Infrared divergent scalar box integrals}
\label{app:int}

\newcommand{\sbb}{\bar s}
\newcommand{\tb}{\bar t}
\newcommand{\sob}{\bar s_1}
\newcommand{\sfb}{\bar s_4}
\newcommand{\li}{{\rm Li_2}}
\def\idelta{i\delta}

In this appendix, we denote the infrared-divergent box integrals with massive
propagators which  emerge in the
calculation of the loop corrections to VBF-induced pentagon diagrams. We do not
list the other scalar loop integrals encountered in our calculation, since they
can be found elsewhere (see, e.g.,
\cite{dn:box,abhs:int,looptools}). 

We obtained the respective soft and collinear singular box diagrams  by 
extracting appropriate expressions from 
Refs.~\cite{Denner:1991qq,Beenakker:1988jr} in the limit of
small quark masses. 
According to Ref.~\cite{ditt:sing}, a relation between
different regularization schemes can be established making use of the genuine
singularity structure of infrared-divergent triangle integrals.
With the help of this property, 
we transformed the divergent four-point integrals from mass
regularization to dimensional regularization.

\begin{figure}[!tb] 
\centerline{ 
\epsfig{figure=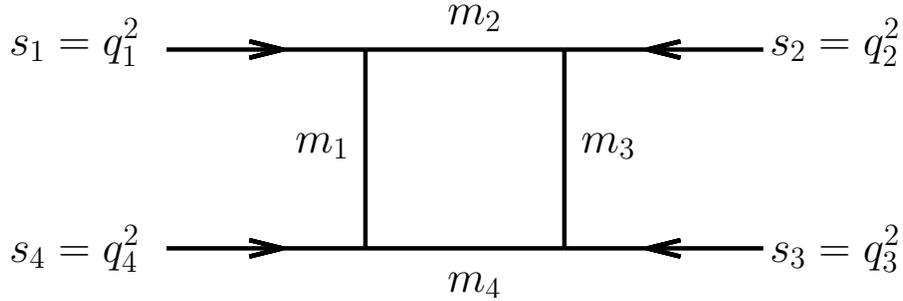,height= 0.17\textheight,clip=}
} 
\caption
{\label{fig:box} 
Momentum and mass assignments for a general scalar box diagram.  
}
\end{figure}

In the following, we refer to a genuine scalar four-point function 
as depicted in Fig.~\ref{fig:box}, 
% 
%%%%%%%%%%%%%%%%%%%%%%%%%%%%%%%%%%%%%%%%%%%%%%%%%%%%%%%%%%%%%%%%%%%%%%
%
% 
%%%%%%%%%%%%%%%%%%%%%%%%%%%%%%%%%%%%%%%%%%%%%%%%%%%%%%%%%%%%%%%%%%%%%%
%
%
\bea
D_0(q_1,q_2,q_3;m_1,m_2,m_3,m_4) 
&=&
\frac{(2\pi\mu)^{4-d}}{(i\pi^2)}
	\int d^d q
	\frac{1}
	{[q^2-m_1^2+\idelta][(q+q_1)^2-m_2^2+\idelta]}        
\non
\\
&&\times
	\frac{1}
	{[(q+q_1+q_2)^2-m_3^2+\idelta][(q+q_1+q_2+q_3)^2-m_4^2+\idelta]}        
\non
\\[2ex]
&\equiv&
I^d_4(s_1,s_2,s_3,s_4;s_{12},s_{23};m_1^2,m_2^2,m_3^2,m_4^2)\,,
\non\\
\eea
where the $q_i$ denote incoming momenta of the external legs and the $m_i$
correspond to the masses of the internally propagating particles. 
The kinematic invariants, $s_i$ and $s_{ij}$, are related to the external 
momenta via  $s_i=q_i^2$ and $s_{ij}=(q_i+q_j)^2$. 
Overlined quantities are defined as $\sbb=s+\idelta$, etc.

\begin{figure}[!tb] 
\centerline{ 
\epsfig{figure=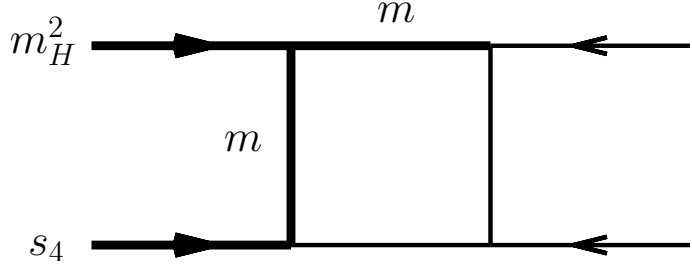,height= 0.15\textheight,clip=}
} 
\caption
{\label{fig:box_coll} 
Momentum and mass assignments for the collinear divergent scalar box diagram
of Eq.~(\ref{eq:coll_box}). Unlabeled thin lines correspond to massless
particles. 
}
\end{figure}

In this notation, the collinear divergent box integral with two equal internal 
and two different external mass scales, which is sketched in
Fig.~\ref{fig:box_coll}, 
% 
%%%%%%%%%%%%%%%%%%%%%%%%%%%%%%%%%%%%%%%%%%%%%%%%%%%%%%%%%%%%%%%%%%%%%%
%
% 
%%%%%%%%%%%%%%%%%%%%%%%%%%%%%%%%%%%%%%%%%%%%%%%%%%%%%%%%%%%%%%%%%%%%%%
%
takes the form 
\bea
\label{eq:coll_box}
\lefteqn{ I^d_4(m_H^2,0,0,s_4;s,t;m^2,m^2,0,0) = \frac{1}{D_1} \Biggl\{ }
\non\\
&&
 \frac{\Gamma(1+\eps)}{\eps}\left(\frac{4\pi\mu^2}{m^2}\right)^{\eps}
 \left[ \ln\left(\frac{m^2-\sfb}{m^2-\sbb}\right) 
       + \ln\left(\frac{m^2}{m^2-\tb}\right)
 \right]
\non\\
&& - \frac{1}{2}\ln^2\left(\frac{-D_1}{m^2(m^2-s_4)}(1+\idelta)\right)
   +\frac{1}{2}\ln^2(-x_{14}k_1)
   +\frac{1}{2}\ln^2\left(-\frac{x_{14}}{k_1}\right)
   -\frac{1}{2}\ln^2(-k_2x_{14})
\non\\
&& -\tpii\theta\left(\frac{(m^2-s)(m^2-s_4)}{D_1}\right) 
   \left[
     \theta\left(\frac{m_H^2-2m^2}{m^2}\right) \ln(x_{14}k_1)
     +\theta\left(\frac{2m^2-m_H^2}{m^2}\right) 
          \ln\left(\frac{x_{14}}{k_1}\right)
   \right.
\non\\
&& \left.
   \hspace*{13em} -\theta\left(\frac{s_4-t}{m^2}\right) \ln(k_2x_{14})
   \right]
\non\\
&& -\tpii\theta\left(\frac{s-m^2}{m^2}\right) 
    \theta\left(\frac{m^2(m^2-s_4)}{D_1}\right)
   \left[
     \ln\left(\frac{D_1(m^2-s)}{m^2(m^2-s_4)^2}(1+\idelta)\right) 
     +\ln\left(\frac{m^2-t}{m^2}-\idelta\right)
   \right]
\non\\
&& -\frac{\pi^2}{6}
   +2\li\left(\frac{\sbb-\sfb}{m^2-\sfb}\right)
   -2\li\left(\frac{-\tb}{m^2-\tb}\right)
   +\li\left(1+\frac{D_1}{m^2(m^2-s_4)}(1+\idelta)\right)
\non\\
&& -\li\left(1+\frac{D_2}{D_1(m^2-s)}-\idelta\frac{m^2(m^2-s_4)}{D_1}\right)
   +\li\left(1+\frac{x_{24}}{k_1}\right) 
   +\eta\left(-x_{24},\frac{1}{k_1}\right)
                               \ln\left(1+\frac{x_{24}}{k_1}\right)
\non\\
&& +\li(1+x_{24}k_1) +\eta(-x_{24},k_1)\ln(1+x_{24}k_1)
   -\li(1+x_{24}k_2) -\eta(-x_{24},k_2)\ln(1+x_{24}k_2)
\non\\
&& +\tpii\theta\left(\frac{s-m^2}{m^2}\right) 
         \theta\left(\frac{m^2(s_4-m^2)}{D_1}\right)
    \left[
      \ln\left(\frac{(m_H^2(m^2-s)+s^2)m^2}{D_1(m^2-s)} 
           -\idelta\frac{m^2(m^2-s)}{D_1}\right)
    \right.
\non\\
&&  \left.
    \hspace*{16em}  +\ln\left(\frac{m^2-t}{m^2}-\idelta\right)
    \right] \Biggr\}\,,
\eea
with
\bea
\eta(a,b) &=& \ln(ab) - \ln(a) - \ln(b),
\non\\
D_1 &=& (s-m^2)(t-m^2)+(s_4-m^2)m^2,
\non\\
D_2 &=& m_H^2(m^2-s)(m^2-t) + m^2[m^2(s+t-s_4)+s(s_4-2t)],
\non
\eea
\bea
\beta &=& \sqrt{1 -4\frac{m^2}{m_H^2} 
               +2\idelta\frac{m^2}{m_H^4}(m_H^2-2m^2)\theta(4m^2-m_H^2) },
\non\\
x_{14} &=& \frac{D_1(m^2-s)}{m^4(m^2-s_4)},
\non\\
x_{24} &=& \frac{D_2}{D_1m^2} -\idelta\frac{(m^2-s)(m^2-s_4)}{D_1},
\non\\
k_1 &=& \frac{2m^2-(1+\beta)m_H^2}{2m^2} 
       +\idelta\frac{2m^2 -(1+\beta)m_H^2}{2\beta m_H^2},
\non\\
k_2 &=& \frac{m^2-s_4}{m^2-t} +\idelta \frac{m^2(t-s_4)}{(m^2-t)^2}\;.
\eea

\begin{figure}
\centerline{ 
\epsfig{figure=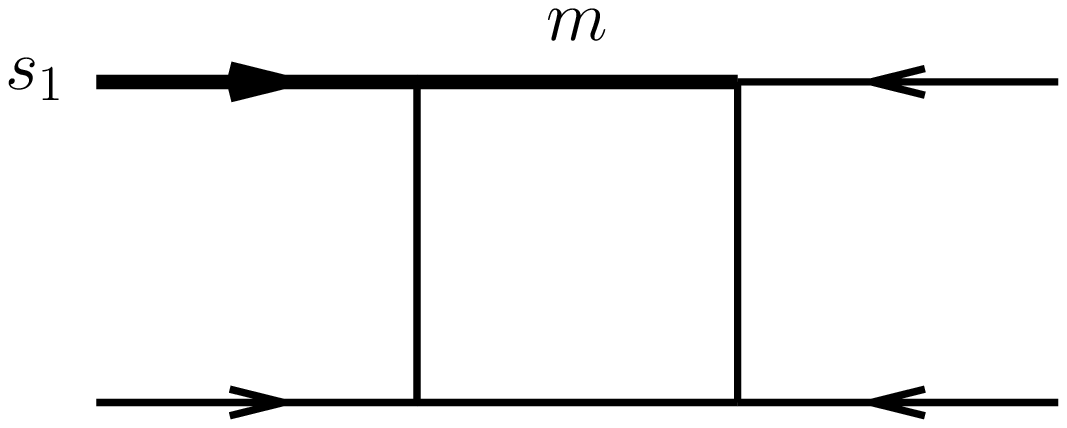,height= 0.15\textheight,clip=}
} 
\caption
{\label{fig:box_soft} 
Momentum and mass assignments for the soft divergent scalar box diagram
of Eq.~(\ref{eq:soft_box}). Unlabeled thin lines correspond to massless
particles. 
}
\end{figure}

The soft divergent box integral with one internal and one external mass scale 
shown in Fig.~\ref{fig:box_soft} 
% 
%%%%%%%%%%%%%%%%%%%%%%%%%%%%%%%%%%%%%%%%%%%%%%%%%%%%%%%%%%%%%%%%%%%%%%
%
% 
%%%%%%%%%%%%%%%%%%%%%%%%%%%%%%%%%%%%%%%%%%%%%%%%%%%%%%%%%%%%%%%%%%%%%%
%
is given by 
\bea
\label{eq:soft_box}
\lefteqn{ I^d_4(s_1,0,0,0;s,t;0,m^2,0,0) = \frac{1}{s(t-m^2)} \Biggl\{ }
\non\\
&&
 \frac{\Gamma(1+\eps)}{\eps^2}\left(\frac{4\pi\mu^2}{-\sbb}\right)^{\eps}
 +\frac{\Gamma(1+\eps)}{\eps}\left(\frac{4\pi\mu^2}{m^2}\right)^{\eps}
    \left[
       \ln\left(\frac{m^2-\sob}{m^2-\tb}\right)
       +\ln\left(\frac{m^2}{m^2-\tb}\right)
    \right]
\non\\
&& -\frac{1}{2}\ln^2\left(-\frac{\sbb}{m^2}\right)
   -2\ln\left(-\frac{\sbb}{m^2}\right)
      \ln\left(\frac{m^2}{m^2-\tb}\right)
   -\ln^2\left(\frac{m^2-\sob}{m^2}\right)
   +\li\left(\frac{\sob-m^2}{\sbb}\right)
\non\\
&& -2\li\left(\frac{\sob-\tb}{m^2-\tb}\right)
   -2\li\left(\frac{-\tb}{m^2-\tb}\right)
   -\ln\left(\frac{\sbb}{\sob-m^2}\right) 
      \ln\left(1-\frac{\sob-m^2}{\sbb}\right)
   -\frac{\pi^2}{2}
\Biggr\}\,.
\non\\
\eea

We would like to note that our results agree with those of Ref.~\cite{abhs:int},
if $\mu^2$ is replaced by $\mu$ in all terms of Eq.~(A.19) 
in~\cite{abhs:int}.

%%%%%%%%%%%%%%%%%%%%%%%%%%%%%%%%%%%%%%%%%%%%%%%%%%%%%%%%%%%%%%%%%

%%%%%%%%%%%%%%%%%%%%%%%%%%%%%%%%%%%%%%%%%%%%%%%%%%%%%%%%%%%%%%%%%

\end{document}